\documentclass[prl,superscriptaddress]{revtex4}

\usepackage[applemac]{inputenc}

\usepackage{graphicx}
\usepackage{amssymb}
\usepackage{amsfonts}

\usepackage[pdftex,bookmarks,colorlinks,breaklinks]{hyperref}  
\hypersetup{linkcolor=red,citecolor=blue,filecolor=pink,urlcolor=magenta} 

\begin{document}

\title{Absorption enhancement in amorphous silicon photonic crystals for thin film photovoltaic solar cells}

\date{\today}

\author{Ounsi El Daif\footnote{Electronic mail: ounsi@eldaif.net}}
\affiliation{Institut des nanotechnologies de Lyon, CNRS, Ecole Centrale et Universit\'{e} de Lyon, Ecully, France}
\affiliation{Institut des nanotechnologies de Lyon, CNRS, INSA et Universit\'{e} de Lyon, Villeurbanne, France}

\author{Emmanuel Drouard}
\affiliation{Institut des nanotechnologies de Lyon, CNRS, Ecole Centrale et Universit\'{e} de Lyon, Ecully, France}

\author{Yeonsang Park}
\affiliation{Institut des nanotechnologies de Lyon, CNRS, INSA et Universit\'{e} de Lyon, Villeurbanne, France}

\author{Alain Fave}
\author{Anne Kaminski}
\author{Mustapha Lemiti}
\affiliation{Institut des nanotechnologies de Lyon, CNRS, INSA et Universit\'{e} de Lyon, Villeurbanne, France}

\author{Xavier Letartre}
\author{Pierre Viktorovitch}
\affiliation{Institut des nanotechnologies de Lyon, CNRS, Ecole Centrale et Universit\'{e} de Lyon, Ecully, France}

\author{Sungmo Ahn}
\author{Heonsu Jeon}
\affiliation{Department of Physics and Astronomy, Seoul National University, Seoul 151-747, Korea}

\author{Christian Seassal}
\affiliation{Institut des nanotechnologies de Lyon, CNRS, Ecole Centrale et Universit\'{e} de Lyon, Ecully, France}

\begin{abstract}
We report on very high enhancement of thin layer's absorption through band-engineering of a photonic crystal structure. We realized amorphous silicon (aSi) photonic crystals, where slow light modes improve absorption efficiency. We show through simulation that an increase of  the absorption by a factor of 1.5 is expected for a film of aSi. The proposal is then validated by an experimental demonstration, showing an important increase of the absorption of a layer of aSi over a spectral range of $0.32-0.76 \mu m$.
\end{abstract}

\maketitle

The efficiency of photovoltaic solar cells is limited, to a large extent, by optical losses. In the case of classical first generation devices, these losses are overcome using anti-reflection coatings and surface texturation \cite{papet06}. In the case of more prospective solar cells using advanced concepts, new materials and thin absorbing layers, many possibilities are explored to control light absorption. Using nanophotonics through surface plasmons \cite{derkacs06} or photonic crystals, it is possible to control further photon capture and absorption, and larger conversion efficiencies may be expected.
In this sense, nanostructured materials have been extensively used for solar energy conversion along a wide variety of conceptual and technological approaches. In particular, wavelength-scale patterning has been efficiently realised by several groups, with an approach usually based on random or periodic structuration of the top surface \cite{tsakalakos07,chhajed08} or patterning of the reflector \cite{zeng05}. 

\begin{figure}[h]
   \centering
   \includegraphics[width=0.5\textwidth]{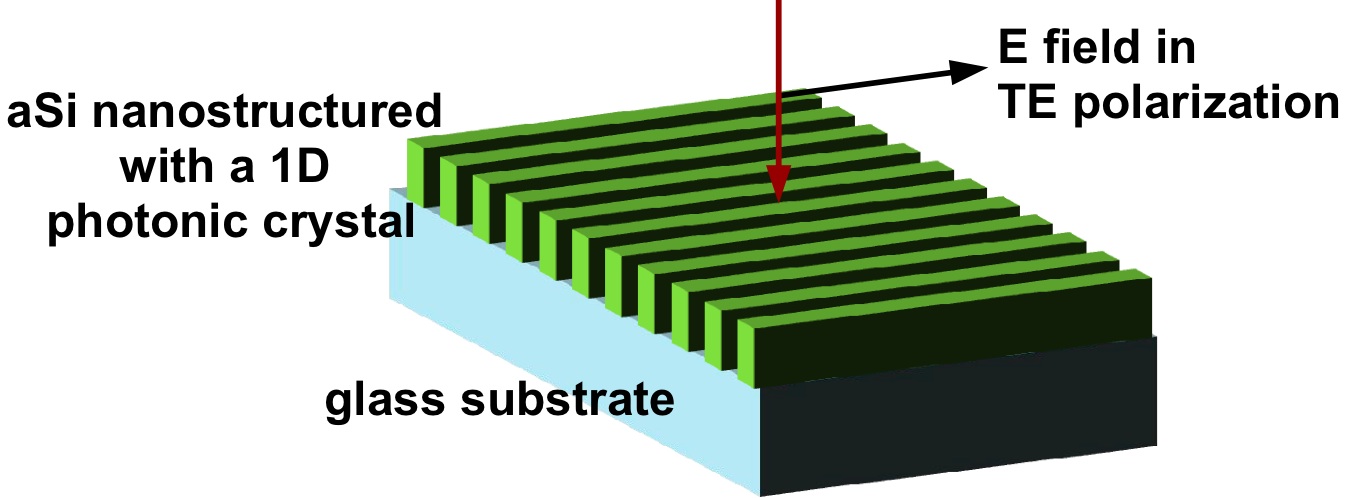}
   \caption{Example of a one-dimensionnal photonic crystal etched into a layer of amorphous silicon.}
   \label{modele}
\end{figure}

We propose to use photonic crystal (PC) slow light modes in order to improve the absorption efficiency of thin absorbing layers, for subsequent embedding in solar cells. It is possible to control the properties of their photonic bands in order to obtain such a slow light mode; through tuning of the parameters of a one or two-dimensional PC, etched into the absorbing layer, in order to optimize light absorption rate. Indeed, when using an absorbing medium for the fabrication of the PC, photons at resonances have higher probabilities to be absorbed thanks to their longer lifetime inside the structure, as was previously shown in ref. \cite{seassal08}. The quality factor of the slow light modes may be tuned in order to set the photon lifetime, and their very flat dispersion characteristics should widen the angular aperture of photon collection. A model structure is shown on Fig. \ref{modele}. Photonic slow Bloch modes have been (and are) studied in various configurations, and allow to produce several kinds of functionalities, such as wide-band reflectors \cite{boutami07}, or high quality factor resonators for laser applications \cite{bakir06,ferrier08}. The concepts we want to demonstrate here is valid in principle for any absorbing material. For demonstration purposes, we will focus in this letter on amorphous silicon (aSi).

This approach is complementary with other technological schemes, specially antireflection coatings. Presently, the best antireflection (AR) coating performances reached on thin solar cells \cite{meier} allowed to reduce light reflection until $\approx 4\%$, on a spectral width of $\approx 150nm$. But the spectral limits imposed by the solar spectrum and the material absorption (usually aSi or crystaline silicon) are wider than that; thus the mode engineering we are proposing here can be done both in order to enhance the absorbing layer performance out of the AR coating spectral limit, and to further lower the $4\%$ reflection reached -through absorption enhancement in the thin absorbing layer.

As a proof of principle, we performed simulations using the Rigorous Coupled-Wave Analysis (RCWA) method on a 100nm thick model layer of amorphous silicon (aSi), focusing on a single Bloch mode. The aim was first to show that a slow light mode can be used with an absorbing medium in order to increase absorption. We showed a $3.5$ times increase of the absorption at resonance in ref. \cite{seassal08}, for a one-dimensional PC for a light field polarized in the direction of the slits (TE polarization, see Fig. \ref{modele}).

In order to allow an experimental demonstration on a wider spectrum, we chose to work with a $1 \mu m$ thick layer of amorphous silicon, likely to be multimode. A one-dimensional photonic crystals optimised for TE polarized light (see Fig. \ref{modele}) was first designed. Therefore, a $1 \mu m$-thick amorphous silicon film on glass was fabricated. Patterning was achieved using holographic lithography, and the resist mask was transferred into a hard mask of $SiO_{2}$. Amorphous silicon was then partially etched using Reactive Ion Etching, with a gas plasma based on $SF_{6}$ and Ar \cite{jamois08}. As amorphous silicon does not show crystalline directions, state-of-the-art etching depth with satisfying sidewalls reaches $\approx 230nm$.

Given this patterning depth we used an optimization loop written with CAMFR \footnote{CAMFR is a language based on RCWA and developed by the Ghent University, see \url{http://camfr.sourceforge.net/}}, in order to find the optimal PC parameters for absorption enhancement. These simulations lead to a maximum absorption for a photonic crystal period of 400nm, an air filling factor of $75\%$ and an etching depth of $200 nm$. The expected total solar light\footnote{The incident light is always taken having the solar spectral intensity distribution.} absorption between $\lambda =0.32 \mu m$ (approximate spectral limit of the solar spectrum) and $0.76\mu m$ (aSi gap) is then $\approx 64\%$, to be compared with $\approx 43\%$ without patterning.

\begin{figure}[h]
   \centering
   \includegraphics[width=0.47\textwidth]{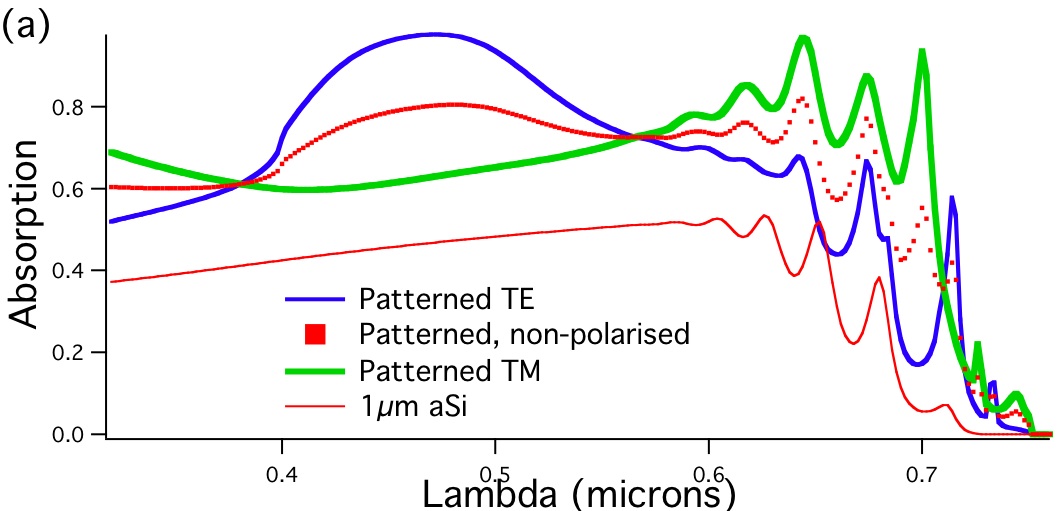}
      \includegraphics[width=0.47\textwidth]{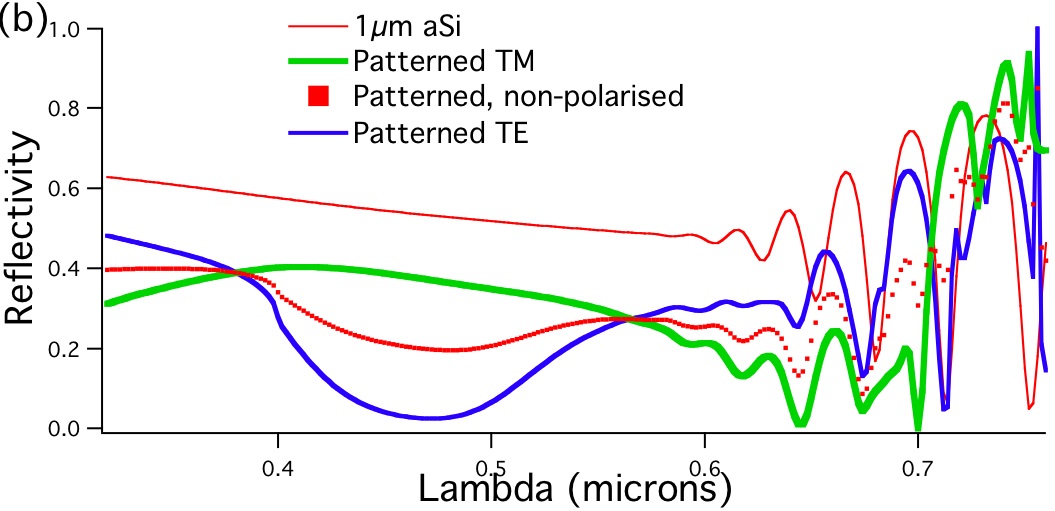}
   \caption{Simulated (a) absorption and (b) resulting reflection of the realised sample for a TE incident light.}
   \label{comparaison}
\end{figure}

Before measuring the spectral response of this optimal sample, we simulated its absorption spectrum, shown on Fig. \ref{comparaison}(a). Refraction index data are taken from ref. \cite{Fontcuberta04}. Fig. \ref{comparaison}(b) shows the corresponding total reflectivity (including specular reflection and diffraction), indeed the latter is the accessible experimental data. Simulated spectra are shown for both TE and TM polarizations, and for the arithmetic mean-value of TE and TM data (taken for non-polarized light). Given the considered thickness of amorphous silicon no light is transmitted for wavelengths $\lambda \lesssim 650 nm$, therefore reflectivity for these wavelengths is a direct information on the absorption. As for wavelengths $650nm < \lambda <730nm$, the aSi layer becomes transparent, yielding then an interference pattern.

We deduce from Fig. \ref{comparaison}(a) an integrated solar absorption of $\approx 65\%$ of the incident TE light for the patterned layer, and $\approx 63\%$ for TM polarization, so $64\%$ on average, as calculated through the optimisation loop. This should be compared to $43\%$ for the non-patterned aSi layer.

In order to characterize the optical properties of this structure, we performed reflectivity experiments integrated over the whole half space around the sample front surface, with an integrating sphere. As mentioned above, given that no light is transmitted below $\approx 630-650nm$ reflectivity $R=1-A$ where A is the absorption in this spectral range. For higher wavelengths, the spectrum is to be compared to the simulated spectrum tendency. 

\begin{figure}[h]
   \centering
   \includegraphics[width=0.5\textwidth]{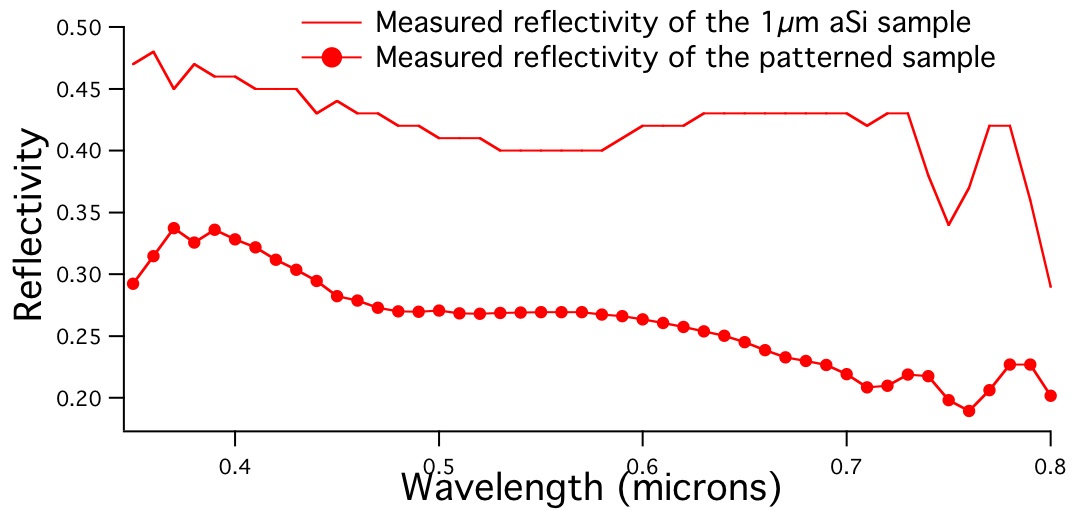}
   \caption{Experimental reflectivity of the patterned sample (round markers) and of the simple $1\mu m$ thick aSi layer (thin line) on glass.}
   \label{refl}
\end{figure}

The reflectivity measurement showed that the patterned film reflectivity is drastically reduced with regards to the non-patterned film reflectivity over the whole considered spectrum, as can be seen on Fig \ref{refl}. This demonstrates, below $630 nm$, an equivalent absorption increase.

These reflectivity measurements exhibit similar tendencies as those observed on the simulations shown on Fig. \ref{comparaison} (b). As the white light source used in the experiment is non-polarized, the patterned sample reflectivity measurement should be compared to the corresponding simulation. A broad peak can be seen for TE light on the simulated curves around 470nm. This peak is attributed to a mode of the PC structure, as it will be discussed below. In the experimental curve, a significant decrease of the reflectivity is observed just below 500nm, we attribute this to a smoothed signature of this mode, observed on the patterned simulated reflectivity curve. At high wavelengths oscillations appear, similar to what is observed on the simulations but smoothed again, and starting at higher wavelength.

\begin{figure}[h]
   \centering
   \includegraphics[width=0.75\textwidth]{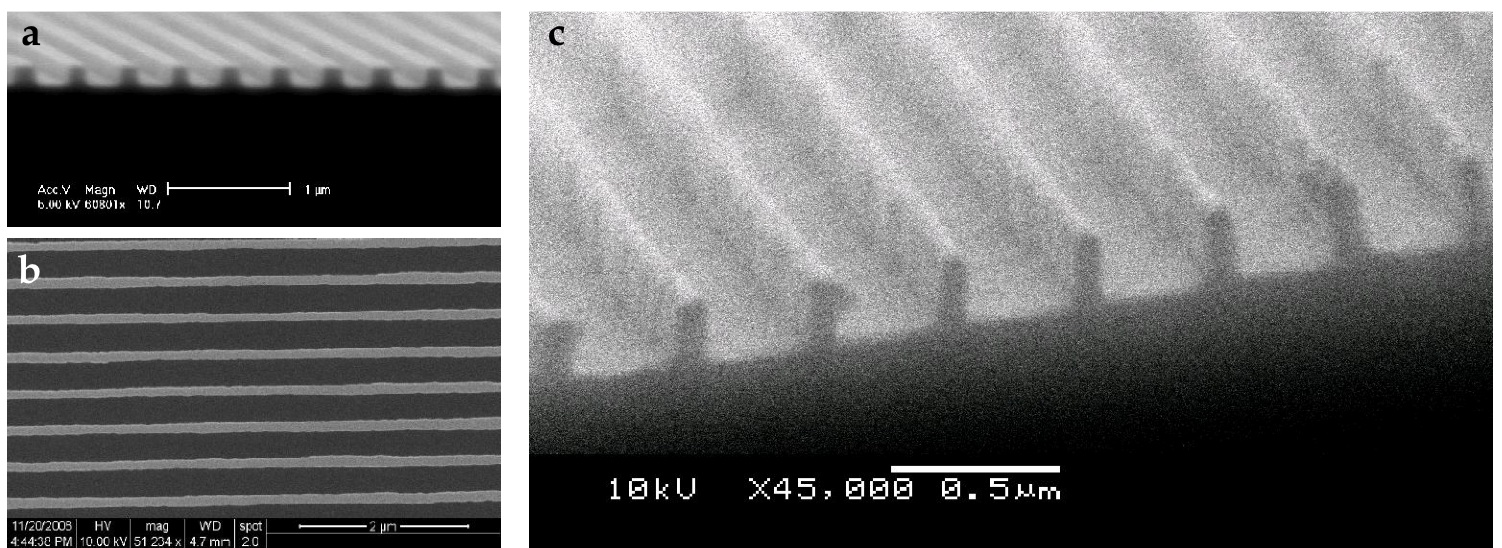}
   \caption{Scanning electron microscope views of patterned aSi: a. Test sample with a metallic film deposited above. b. aSi etched surface seen from above. c. Profile view, the aSi has been etched on $\approx 200 \pm 15 nm$.}
   \label{meb}
\end{figure}

One reason for the experiment/simulation disagreements may be in the amorphous silicon optical index difference between the real and simulated samples. Moreover we also attribute these differences to technological uncertainties; The topographical profile of the sample is shown on Fig.\ref{meb}(a). The actual aSi sample is shown on Fig.\ref{meb}(b) and (c). One can see on Fig.\ref{meb}(c) that, for the etching depth chosen, the sidewalls are quite vertical, but the patterns section is not perfectly rectangular. The lines seen on both Fig.\ref{meb}(b) and (c) are not perfectly strait, yielding an unavoidable disorder (due both to holographic lithography, and to etching of an amorphous material) this may smooth the measured electromagnetic spectra.

\begin{figure}[h]
   \centering
   \includegraphics[width=0.4\textwidth]{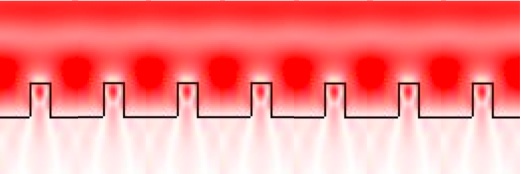}
   \caption{(color online) Mapping of the electromagnetic field intensity in the patterned structure at a wavelength $\lambda = 480 nm$.}
   \label{map}
\end{figure}

In order to confirm and to understand better the role of the PC structure, we performed a mapping of the simulated electromagnetic field in the studied structure, for a TE polarized field at a wavelength of 480nm, where the absorption mode is centered. One can see on Fig. \ref{map} that maxima of electromagnetic field intensity distribution are confined into the mesas lines, therefore showing that the PC is yielding most of the absorption.

In conclusion we demonstrate here both theoretically and experimentally a method to drastically improve the absorption of thin amorphous silicon layers for photovoltaic applications. This method is applicable to any absorbing material, such as crystalline silicon or III-V semiconductors, given an adaptation of the PC parameters to the complex refraction index of the material. Given the controllability of the spectral properties of the absorption enhancement, we believe this method could also be used to improve the sensitivity of detectors on given spectral bands \cite{bandiera:193103}. We are currently performing simulation and design of entire solar cells based on this method: the combination of slow light modes and of efficient AR coatings should further increase absorption efficiency. It is nevertheless needed to evaluate the photoelectrical losses which may be induced by the etching process. Current simulations performed in our group show that the latter are more than compensated for by the improved absorption yield \cite{seassal08}.

\

We acknowledge funding from the French National Research Agency (ANR) Solar Photovoltaic program (SPARCS project). This work was partly performed in the frame of the French-Korean International Associated Lab "Center for Photonics and Nanostructure". We thank Dr. C. Jamois for helpful discussions, and the NanoLyon technology performed, in particular P. Cr\'{e}millieu and Dr. R. Mazurczyk.

\end{document}